\documentclass[,final
  ]
  {aipproc}

\layoutstyle{6x9}

\usepackage{graphicx}

\begin{document}

\title[Eclipsing and transiting sub-stellar and planetary companions
to white dwarfs]{Detection limits for close eclipsing and transiting
  sub-stellar and planetary companions to white dwarfs in the WASP
  survey}


\author{F. Faedi}{
  address={Astrophysics Research Centre, School of Mathematics and Physics, Queens University, Belfast, BT7 1NN, U.K.}
,altaddress={Department of Physics and Astronomy, University of Leicester, Leicester, LE1 7RH, U.K.}
}

\author{R.~G.~West}{
  address={Department of Physics and Astronomy, University of Leicester, Leicester, LE1 7RH, U.K.}
}

\author{M.~R.~Burleigh}{
  address={Department of Physics and Astronomy, University of Leicester, Leicester, LE1 7RH, U.K.}
}

\author{M.~R.~Goad}{
 address={Department of Physics and Astronomy, University of Leicester, Leicester, LE1 7RH, U.K.}
}

\author{L.~Hebb}{
  address={Department of Physics and Astronomy, Vanderbilt University, Nashville, TN 37235, U.S.A.}
,altaddress={School of Physics and Astronomy, University of St. Andrews, North Haugh, Fife, KY16 9SS, U.K.}
}

\begin{abstract}

  We used photometric data from the WASP
  (Wide-Angle Search for Planets) survey to explore the
  possibility of detecting eclipses and transit signals of brown
  dwarfs, gas giants and terrestrial companions in close orbit around
  white dwarfs. We performed extensive Monte Carlo simulations and we
  found that for Gaussian random noise WASP is sensitive to companions
  as small as the Moon orbiting a $V\sim$12 white dwarf. For fainter
  stars WASP is sensitive to increasingly larger bodies. Our
  sensitivity drops in the presence of co-variant noise structure in
  the data, nevertheless Earth-size bodies remain readily detectable
  in relatively low S/N data. We searched for eclipses and transit
  signals in a sample of 194 white dwarfs in the WASP archive however,
  no evidence for companions was found. We used our results to place
  tentative upper limits to the frequency of such systems. While we
  can only place weak limits on the likely frequency of Earth-sized or
  smaller companions; brown dwarfs and gas giants (radius$\simeq$
  R$_{jup}$) with periods $\leq$0.2 days must certainly be rare
  ($<10\%$). More stringent constraints requires significantly larger
  white dwarf samples, higher observing cadence and continuous
  coverage. The short duration of eclipses and transits of white
  dwarfs compared to the cadence of WASP observations appears to be
  one of the main factors limiting the detection rate in a survey
  optimised for planetary transits of main sequence stars.

\end{abstract}

\maketitle


\section{Introduction}

\noindent
The transit technique involves searching for periodic dips in stellar
light-curves due to the orbital revolution of a transiting body,
blocking a fraction of the stellar light. For a given planetary
radius, the transit depth ($\delta$) is proportional to
(R$_{pl}$/R$_*)^2$. Therefore, planets orbiting solartype stars have
extremely shallow eclipses, blocking $\sim$1\% of the light for a
giant planet and $\sim$0.01\% of the light for an Earth-sized planet.
Current ground-based wide-field surveys can achieve the necessary
photometric accuracy of better than 1\%, only for the brightest stars
($V\sim$9-12 in the case of WASP), so the bulk of the planets
discovered by transit surveys around main-sequence stars have radii in
the range R$_{pl}\sim$0.8$-$1.8 R$_{jup}$. A strong advantage over
main sequence star primaries is offered by white dwarf stars. White
dwarfs (WD) are compact degenerate objects, with approximately the
same radius as the Earth, and represent the final stage of evolution
of main-sequence stars with masses~$<$~8M$_{\odot}$ (i.e. $\sim$97\% of
all stars in our Galaxy). Any sub-stellar or gas giant companion in
orbit around a white dwarf will completely eclipse it, while bodies as
small as the Moon will have relatively large transit depths
($\sim$~3\%), with the only caveat being that it remains unclear as to
whether any such systems
survive beyond the latter stages of stellar evolution.\\

\noindent
Sub-stellar companions to white dwarfs are rare (\citealt{Farihi05} using 2MASS
estimated that $<0.5\%$ of WDs have L dwarf companions). At the time
of writing only three wide white dwarf + brown dwarf (WD+BD) systems
have been spectroscopically confirmed, GD\,165 \citep{Becklin88},
PHL5038 \citep{Steele09}, and LSPM~$1459+0857$\,AB \citep{Day-Jones}
and two detached, non-eclipsing, short-period WD+BD systems are
currently known, WD$0137-349$ (\citealt{Maxted06}, \citealt{wd0137b},
P$\approx$116mins), and GD1400 (\citealt{Farihi04},
\citealt{Dobbie05}, \citealt{Burleigh10}, P$\approx$9.9h). The latter,
is currently the lowest mass ($\sim$50M$_{jup}$) object known to have
survived CE evolution. Although infrared surveys such as UKIDSS, VISTA
and WISE, and observatories such as Spitzer hope to reveal many more
such binaries, they remain difficult to identify either as infra-red
excesses or through radial velocity measurements. In addition the
detection of a significant number of eclipsing WD+BD binary systems
might help uncover the hypothesised population of `old' cataclysmic
variables (CVs) in which the companion has been reduced to a
sub-stellar mass (e.g. \citealt{Patterson98}; \citealt{Patterson05};
\citealt{Littlefair03}). These systems are undetectable as X-ray
sources and difficult to identify in optical and infra-red surveys.
\citet{Littlefair06} confirmed the first such system through eclipse
measurements, while \citet{Littlefair07} showed that another eclipsing
CV, SDSS~J$150722.30+523039.8$, was formed directly from a detached
WD+BD binary.\\

\noindent
Several theoretical studies discuss post-main sequence evolution of
planetary systems and show that planetary survival is not beyond
possibility (\citealt{Duncan98}; \citealt{Debes02};
\citealt{Burleigh02}; and \citealt{Villaver07}). Radial velocity
observations of red giants indicate that planets in orbits beyond the
red giant's envelope can survive stellar evolution to that stage (see
\citealt{Frink02}; \citealt{Hatzes05}, \citealt{Sato03}). Moreover,
\citet{Silvotti07} reported the detection of a $\sim$3M$_{jup}$ planet
orbiting an extreme horizontal branch star, and \citet{Mullally08}
found convincing evidence of a 2M$_{jup}$ planet in a 4.5 year orbit
around a pulsating WD. Furthermore, \citet{Beuermann10} reported the
detection of two planetary companions (M$_c$=6.9M$_{Jup}$ and
M$_d$=2.2M$_{Jup}$) in the post common envelope binary NN Ser (ab) via
measurements of a light-travel-time effect superposed on the
linear ephemeris of the binary; showing that planets do survive stellar evolution.\\

\noindent
Short-period rocky companions to white dwarfs may seem less likely.
\citet{Villaver07} suggested that planets in orbit within the reach of
the AGB envelope will either evaporate or in rare cases, more massive
bodies may accrete mass and become close companions to the star.
Planets in wide orbits that escape engulfment by the red giant or
asymptotic giant will move outwards to conserve angular momentum (as
described by Jeans 1924). \citet{Duncan98} found that for WD
progenitors experiencing substantial mass loss during the AGB phase,
planetary orbits become unstable on timescales of $\leq$10$^8$~year.
\citet{Debes02} found that the mass loss is sufficient to destabilise
planetary systems of two or more planets and that the most likely
result is that one planet would be scattered into an inner orbit
(occupied, before the RGB phase, by a `now evaporated' inner planet),
while the other would either be boosted into a larger orbit, or
ejected from the system altogether.\\

\noindent
The above scenario provides a plausible explanation for the recent
detection of silicate-rich dust discs around a growing number of white
dwarfs at orbital radii up to $\sim$1R$_{\odot}$ (e.g.
\citealt{Reach05}; \citealt{Farihi07}, 2008; \citealt{Jura03}).
\citet{Jura03} suggests that the formation of dust discs around white
dwarfs is most probably due to the tidal disruption of an asteroid or
larger body which has strayed too close to the parent star.
(\citealt{Jura09}) suggest that the disc around GD362 originated from
the tidal destruction of a single massive body such
as Callisto or Mars.\\

\noindent
The detection of short period sub-stellar and planetary mass
companions to white dwarfs, will open an exciting chapter in the study
of exoplanet evolution, constraining theoretical models of CE
evolution and helping us to understand the ultimate fate of hot
Jupiter systems as well as the fate of our own solar system in the
post main-sequence phase. Here we present some of the results of our
study which investigated the detection limits for transiting
sub-stellar and terrestrial companions in close orbits around white
dwarfs (for more details see \citealt{Faedi10}).\\

\noindent
In $\S$2 we discuss the characteristics of the transit signals, the
parameter space investigated and our detection method. In $\S$3 we
analysed a sample of 194 WDs in the WASP archive. Finally in $\S$4 we
discuss our conclusions.

\section{Characteristics of the transit signal}
A transit signal is described by its {\em duration}, its {\em depth}
and its {\em shape}. Extra-solar planets transiting main-sequence
stars show signals characterised by an ingress, a flat bottom and an
egress, with a typical durations of 2-3 hours and depths of about 1\%
(see for example \citealt{Cameron07}; \citealt{Simpson10};
\citealt{Barros10}; and Faedi et al. 2010). We modelled the
synthetic dataset assuming circular orbits and fixed stellar
parameters. We considered a typical 1~Gyr old carbon-core white dwarf
of mass $M_*=0.6$M$_{\odot}$\ and radius R$_*=0.013$R$_{\odot}$. We
explored the detectability of planetary transits across the
two-dimensional parameter space defined by the orbital period and the
planet radius. Our simulations cover companions $\sim0.3{\rm
  R}_{\oplus}<{\rm R}_{pl}<12{\rm R}_{\oplus}$, and orbital periods in
the range $P\sim 2$ hours to 15 days (equivalent to orbital distances
between $a\sim 0.003$ and 0.1\,AU). We chose the minimum orbital
period to yield an orbital separation close to the Roche radius of the
WD, and the maximum period in order to have a reasonable chance of
detecting five or more transits in a typical WASP season of 150 day.

\begin{figure} 
 \includegraphics[width=0.55\textwidth]{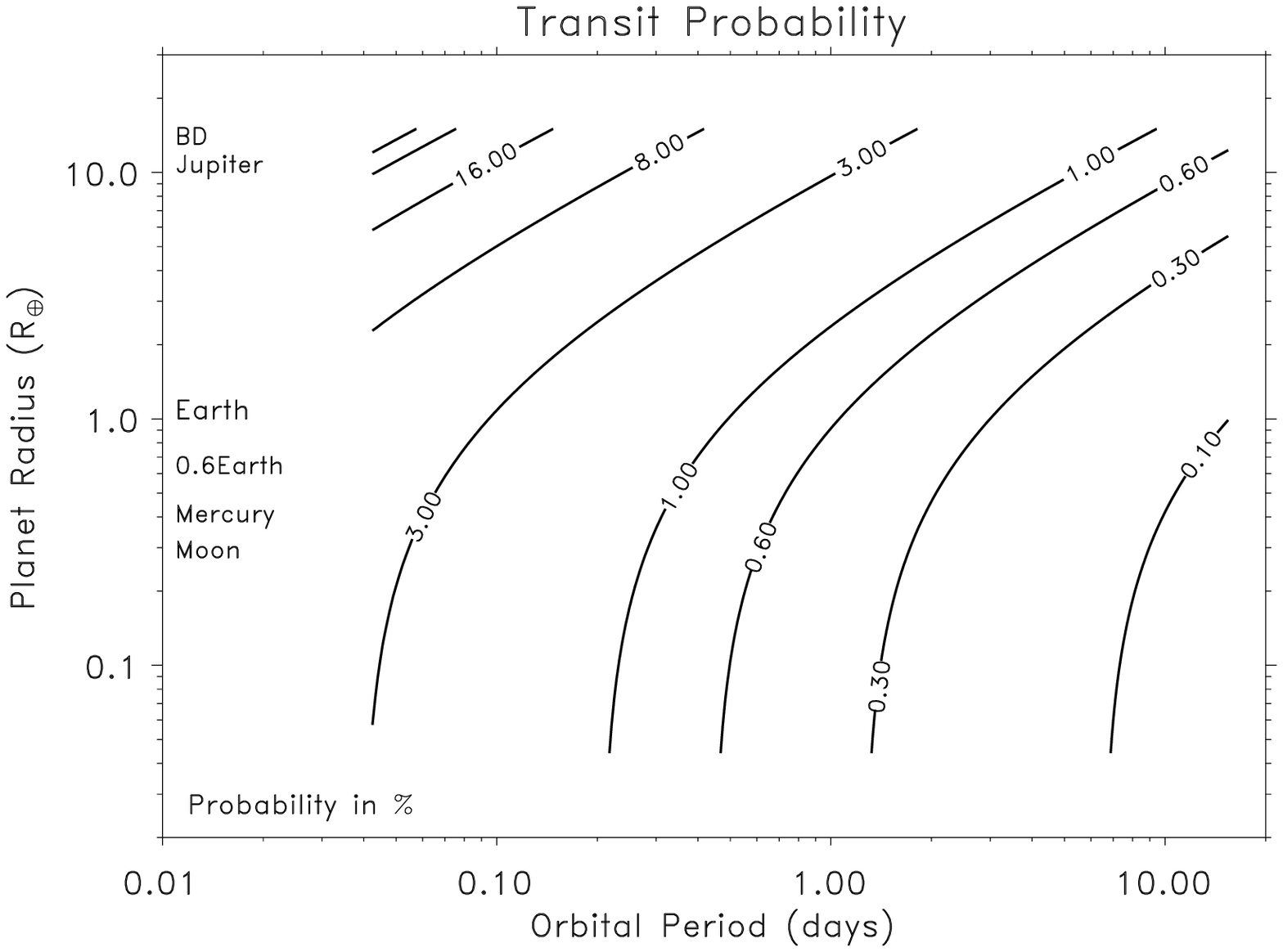}
\includegraphics[width=0.55\textwidth]{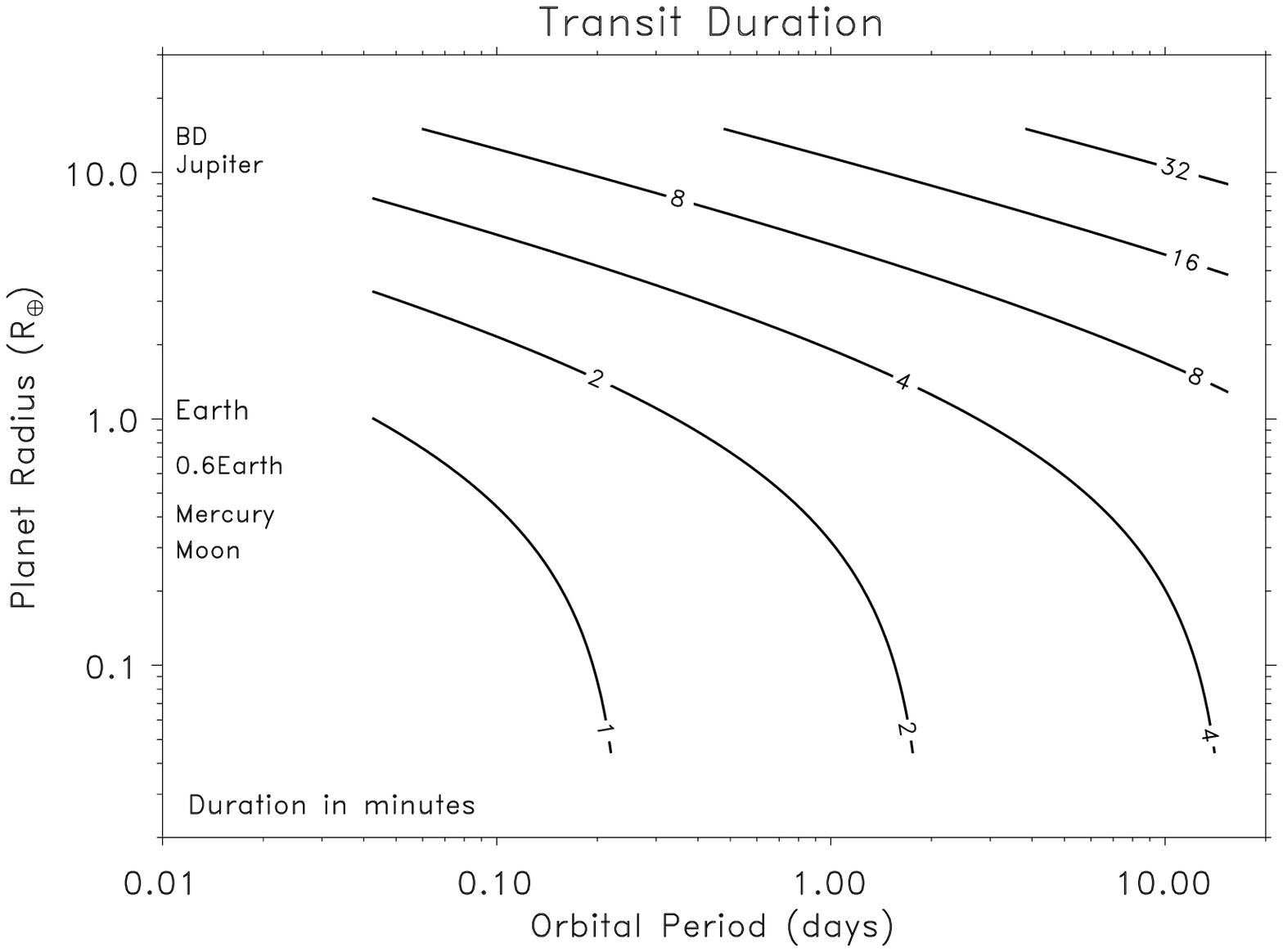}
\caption{Contours of constant transit
    probability (left), and duration (right) in the parameter
    space defined by orbital period and planetary radius. The transit 
    probability is expressed in percentage values. The transit
    duration is expressed in minutes.}
  \label{fig1} 
\end{figure}
In Figure \ref{fig1} we show the probability that a given system will
transit, and the duration of such transit across the parameter space
defined above. It is evident from these diagrams that the signatures
of transits of white dwarfs by typical planet-sized bodies will be
rather different than those seen for typical transiting hot Jupiters.
In particular the transit duration is much shorter, from $\sim$1-30
min for companions with sizes ranging from Moon-size to Jupiter-size,
compared to 2-3 hours for a typical hot Jupiter. In addition, Figure
\ref{fig2}, left-panel, shows that the transit depths are much larger,
from around 3\% for a Moon-sized to 100\% for any companion larger
than the Earth, compared to $\sim$1\% for a hot Jupiter.

\subsection{Synthetic WASP light-curves}

The synthetic light-curves were generated using the time sampling of a
typical WASP survey field, and with statistical S/N representative of
magnitude spanning the range of brightness of WDs in the WASP survey.
The corresponding photometric accuracy of WASP over this range is
$\sim$1\% to 10$\%$. Because WASP data show residual covariant-noise
structure we have tested the transit recovery rate in the case of both
uncorrelated ``white'' noise and correlated ``red'' noise. To cover
the orbital period-planet radius parameter space we selected seven
trial periods spaced approximately logarithmically ($P=0.08$, 0.22,
0.87, 1.56, 3.57, 8.30 and 14.72 days), and five planet radii
R$_{pl}=10.0$, 1.0, 0.6, 0.34 and 0.27~R$_{\oplus}$. We modelled the
set of synthetic light-curves by injecting fake transit signals into
phase-folded light-curves at each trial period with a random transit
epoch $t_0$ in the range $0<t_0<P$. Because in the case of a WD host
star considered here, the ingress and egress duration is typically
short compared to cadence of the WASP survey (8-10 minutes), we
ignored the detailed shape of the ingress and egress phases and
modelled the transit signatures as simple box-like profiles. Figure
\ref{fig2}, right-panel shows two examples of our simulated transit
light-curves. The top panel shows the synthetic light-curve of an
hypothetical eclipsing WD$+$BD binary system with an orbital period of
$P=116$~mins, similar to WD$0137-349$ (a non-eclipsing system,
\citealt{Maxted06}). The lower panel shows the simulated transit
light-curve for a rocky body of radius $1.2$R$_{\oplus}$ in a
5\,hr orbit.\\

\subsection{Detecting transit signals}

To recover the transit signals from the synthetic light-curves we used
an implementation of the box-least-squares (BLS) algorithm
\citep{Kovacs02} commonly used to detect transits of main sequence
stars. To ensure that the BLS search was sensitive across the expected
range of transit durations, we chose to search a grid of box widths
$W_{b}=\{1, 2, 4, 8, 16, 32\}$ minutes, covering the range in transit
durations over most of our parameter space (Figure~\ref{fig1}). In
addition, we used an optimised version of the BLS code which best
accounted for the shape, duration and depth of the signals
investigated in this work \citep{Faedi10}.

\begin{figure} 
 \includegraphics[width=0.55\textwidth]{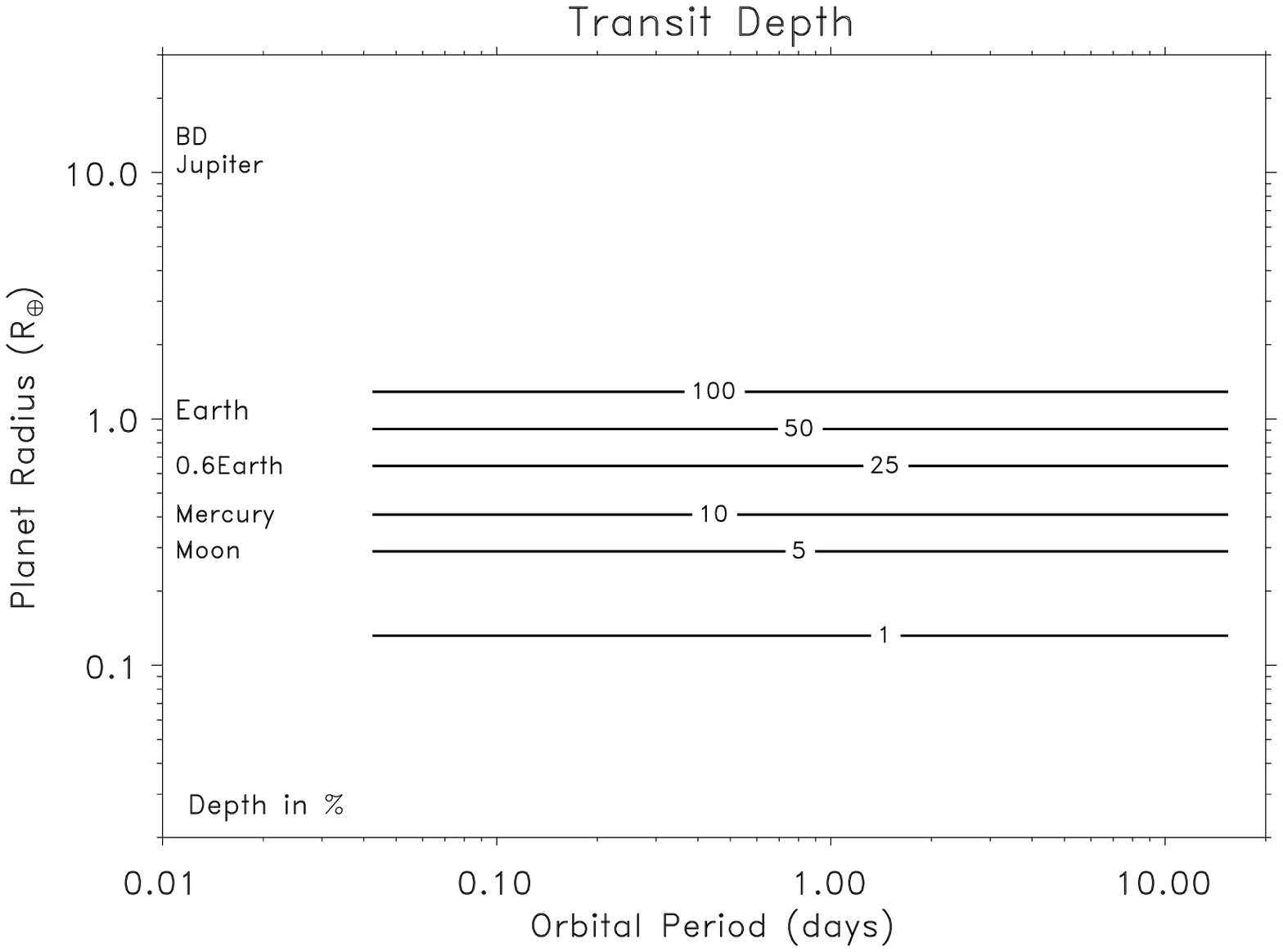}
\includegraphics[width=0.55\textwidth]{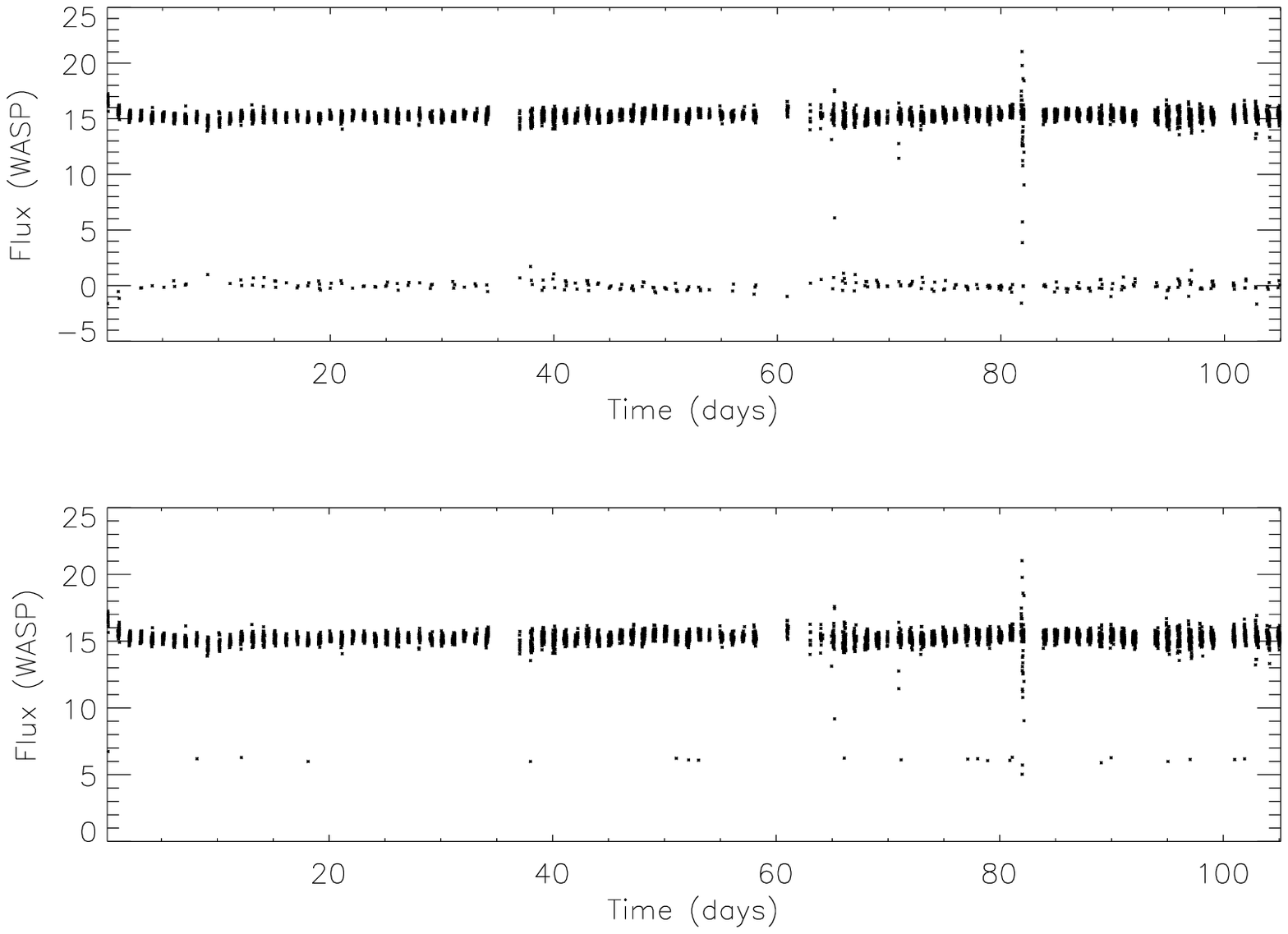}
\caption{Left-panel: contours of constant transit
    depth; right-panel: examples of synthetic light-curves. Top, an eclipsing BD in orbit
    with 2\,hr period. Bottom, a $1.2$R$_{\oplus}$ companion in 5\,hr orbit.}
  \label{fig2} 
\end{figure}
\noindent
We used the Signal Detection Efficiency ($SDE$) metric defined in
\citet{Kovacs02} to assess the likely significance of a peak in a BLS
periodogram. We evaluated $SDE$ as follows:
\[
SDE=\frac{S_{peak} -
  \bar{S}}{\sigma_{S}}
\]
where $S_{peak}$ is the height of the peak, and $\bar{S}$ and
$\sigma_S$ are measures of the mean level and scatter in the
noise continuum of the periodogram. A detection is represented by the
highest peak in the BLS power spectrum. We regard as a match any trial
in which the most significant detected period is within 1\% of being
an integer fraction or multiple from $1/5\times$ to $5\times$ the
injected transit signal. Details of the algorithm false alarm
probability can be found in \citet{Faedi10}. The results of our
simulations are illustrated in Figure \ref{fig2}, left-panel in the
case of a $V\sim$12 magnitude WD for light-curves with red noise. It
is evident from Figure \ref{fig2}; Figure~6 and Table~1,~2, and~3 from
\citet{Faedi10}, that transiting companions are essentially
undetectable at our longest trial periods (8.30 and 14.72 days) in a
WASP-like survey; the transits are too short in duration and too
infrequent to be adequately sampled. In addition, we found that for
idealised photon-noise-limited cases, objects as small as Mercury
could be detected to periods of around 1.5\,d, and the Moon for
periods less than 1\,d. Once red noise is added, Moon-sized companions
become almost undetectable. However, for companions around
1R$_{\oplus}$ and larger there is a good chance of detection out to
periods of around 4 days.
\\
Our key conclusion from these simulations is that for the case of
transits of white dwarfs the degree of photometric precision delivered
by a survey is of somewhat secondary importance compared to a high
cadence and continuous coverage. For planet-sized bodies individual
transits will be quite deep and readily detectable in data of moderate
photometric quality, however it is the short duration of the transits
that is the main factor limiting the transit detection rate in surveys
optimised for main sequence stars.

\section{Searching for transit signals in WASP survey data}

Encouraged by the results of our simulations we selected a sample of
194 WDs (with $V<15$) which have been routinely monitored by WASP
through the 2004 to 2008 observing seasons, and performed a systematic
search for eclipsing and transiting sub-stellar and planetary
companions. We selected the sample by cross-correlating the WASP
archive with the McCook~\&~Sion catalogue \citep{WDcat}. In addition
to our authomated search, we have inspected each of the individual
light-curves by eye. In both searches we found no evidence for any
transiting and eclipsing companions. We have used this null result
together with the results of simulations to estimate an upper-limit to
the frequency of such close companions for the sample of WDs
considered in this study.

\begin{figure} 
\includegraphics[width=0.55\textwidth]{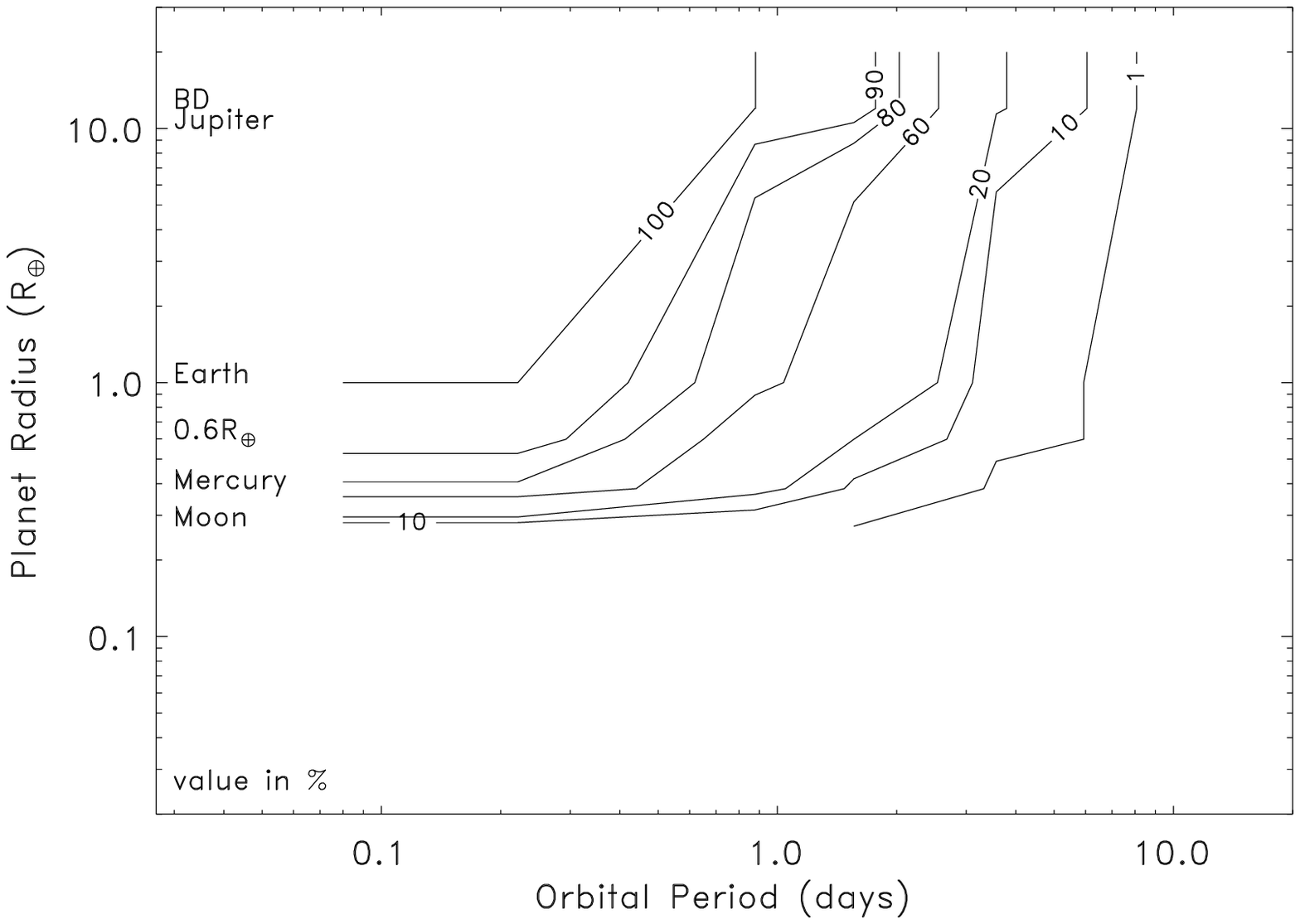}
\includegraphics[width=0.55\textwidth]{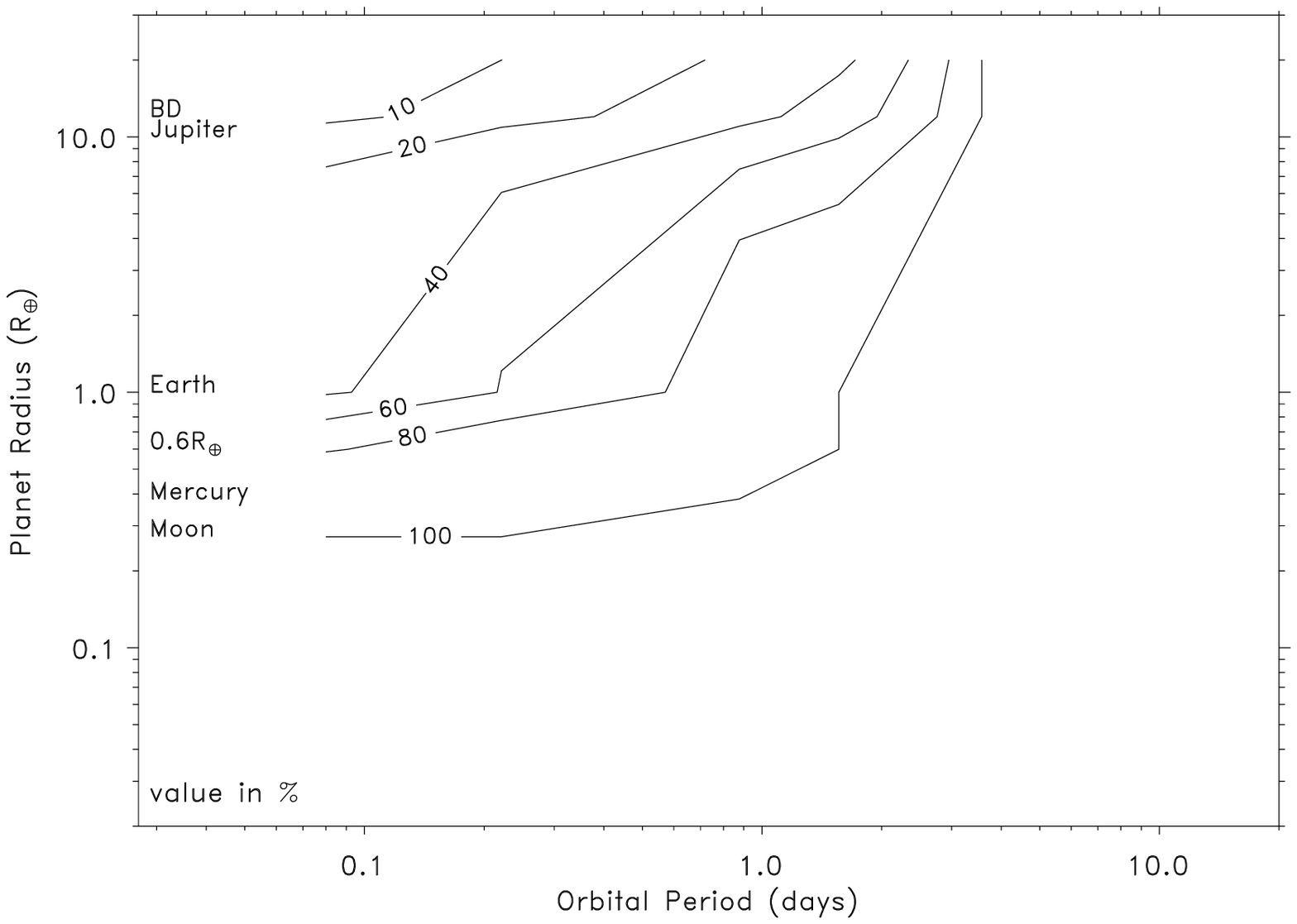}
\caption{Left-panel: recovery rate for simulated transit signals
injected into synthetic light-curves of a white dwarf of magnitude
$V\simeq12$; right-panel: upper-limits on companion frequency (95\%)  folding-in the
    detectability of transiting systems in a WASP-like survey. In both panels values
    are expressed in percent.}
  \label{fig3} 
\end{figure}

In order to estimate an upper limit to the frequency of close
sub-stellar and planetary companions to white dwarfs, we used the
detection limits derived from our simulations and the results obtained
from the analysis of the sample of 194 white dwarfs. Although our
complete sample numbers $N=194$ stars, only a fraction $p_{
  tr}(R_{pl}, P)$ will exhibit a transit, and of those only a fraction
$p_{det}(R_{pl}, P)$ would be detectable in a WASP-like survey. Both
these factors act to reduce the total number of transiting companions
detected in the survey, or in the case of a null result will tend to
weaken the constraints that can be placed on true companion frequency
by such a survey. We incorporate these factors and we modified our
effective sample size as $N'=N\times p_{tr}(R_{pl}, P)\times p_{
  det}(R_{pl}, P)$. We combined the magnitude-specific $p_{det}$ maps
obtained from our simulations for WDs of magnitude $V\sim12$,~13,~15
into a single map by interpolating/extrapolating according to the
magnitude of each object in our sample and combining these to form an
averaged map which can be folded in to our calculation of the
upper-limits. The resulting limits corresponding to the 95\% of the
integrated probability, are shown in the right panel of
Figure~\ref{fig3}. Our results show that for rocky bodies smaller than
the size of Mercury no useful upper limits to the frequency of
companions to white dwarfs can be found, and that for Earth-sized
companions only weak constraints can be imposed. However, it does
suggest that objects the size of BDs or gas giants with
orbital periods P$<0.1-0.2$~days must be relatively rare (upper limit
of $\sim$10\%).

\section{Conclusion}
We have investigated the detection limits for sub-stellar and
planetary companions to white dwarfs using in the WASP survey. We
found that Mercury-sized bodies at small orbital radii can be detected
with good photometric data even in the presence of red noise. For
smaller bodies red noise in the light-curves becomes increasingly
problematic, while for larger orbital periods, the absence of
significant numbers of in-transit points, significantly decreases our
detection sensitivity. Application of our modified BLS algorithm to
search for companions to WDs in our sample of 194 stars available in
the WASP archive, did not reveal any eclipsing or transiting
sub-stellar or planetary companions. We have used our results, to
place upper limits to the frequency of sub-stellar and planetary
companions to WDs. While no useful limits can be placed on the
frequency of Mercury-sized or smaller companions, slightly stronger
constraints can be placed on the frequency of BDs and gas giants with
periods $<0.1-0.2$days, which must certainly be relatively rare
($<10\%$). More stringent constraints would requires significantly
larger WD samples. Our key conclusion from simulations and analysis,
using WASP data, suggests that photometric precision is of secondary
importance compared to a high cadence and continuous coverage. The
short duration of eclipses and transits of WDs compared to the WASP
observing cadence, appears to be the main factor limiting the transit
detection rate in a survey optimised for planetary transits of main
sequence stars. Future surveys such as Pan-STARRS and LSST will be
capable of detecting tens of thousands of WDs. However, we emphasise
the importance of high cadence and long baseline observation when
attempting to detect the signature of close, eclipsing and transiting
sub-stellar and planetary companions to WDs. Space missions such as
{\it COROT}, {\it Kepler} (see \citealt{DiStefano}) and, especially,
{\it PLATO} may therefore be better suited to a survey of white dwarfs
as they deliver uninterrupted coverage at high cadence and exquisite
photometric precision ($\sim 10^{-4}-10^{-5}$) and could at least in
principle detect the transits of asteroid-sized bodies across a white
dwarf.


  



\bibliographystyle{aipproc}   

\bibliography{Faedi.bib}


\end{document}